\documentclass[fleqn,usenatbib]{mnras}

\usepackage{mathptmx}

\usepackage[T1]{fontenc}

\DeclareRobustCommand{\VAN}[3]{#2}
\let\VANthebibliography\thebibliography
\def\thebibliography{\DeclareRobustCommand{\VAN}[3]{##3}\VANthebibliography}

\usepackage{graphicx}	
\usepackage{amsmath}	
\usepackage{amssymb}	
\usepackage{xspace} 
\usepackage{pdflscape}  

\newcommand{\degree}{^\circ}

\newcommand{\angstrom}{Å}

\usepackage{scalerel}
\usepackage{tikz}
\usetikzlibrary{svg.path}

\definecolor{orcidlogocol}{HTML}{A6CE39}
\tikzset{
  orcidlogo/.pic={
    \fill[orcidlogocol] svg{M256,128c0,70.7-57.3,128-128,128C57.3,256,0,198.7,0,128C0,57.3,57.3,0,128,0C198.7,0,256,57.3,256,128z};
    \fill[white] svg{M86.3,186.2H70.9V79.1h15.4v48.4V186.2z}
                 svg{M108.9,79.1h41.6c39.6,0,57,28.3,57,53.6c0,27.5-21.5,53.6-56.8,53.6h-41.8V79.1z M124.3,172.4h24.5c34.9,0,42.9-26.5,42.9-39.7c0-21.5-13.7-39.7-43.7-39.7h-23.7V172.4z}
                 svg{M88.7,56.8c0,5.5-4.5,10.1-10.1,10.1c-5.6,0-10.1-4.6-10.1-10.1c0-5.6,4.5-10.1,10.1-10.1C84.2,46.7,88.7,51.3,88.7,56.8z};
  }
}

\newcommand\orcidicon[1]{\href{https://orcid.org/#1}{\mbox{\scalerel*{
\begin{tikzpicture}[yscale=-1,transform shape]
\pic{orcidlogo};
\end{tikzpicture}
}{|}}}}

\usepackage{hyperref} 

\title[Fireball observation using the ALIS\_4D]{Observations of the 2023 February 27 fireball in northern Sweden using the auroral imaging system ALIS\_4D}
\author[G. Borderes-Motta et al.]{Gabriel Borderes-Motta\orcidicon{0000-0002-4680-8414},$^{1,2}$\thanks{E-mail: gabriel@asu.cas.cz}\
Daniel Kastinen\orcidicon{0000-0002-6371-1016},$^{1}$\
Tima Sergienko\orcidicon{0000-0003-4515-2174},$^{1}$\
\newauthor
Urban Brändström\orcidicon{0000-0003-4273-1947},$^{1}$\
Johan Kero\orcidicon{0000-0002-2177-6751},$^{1}$\
Jaakko Visuri,$^{3}$\
Maria Gritsevich\orcidicon{0000-0003-4268-6277},$^{1,3,4}$\
\newauthor
Jarmo Moilanen\orcidicon{0000-0002-6410-3709},$^{3}$\
Daniela Cardozo Mour\~{a}o\orcidicon{0000-0001-9555-8143}$^{5}$\ and
\newauthor
Barbara Celi Braga Camargo\orcidicon{0000-0003-3937-7297}$^{5,6}$\
\\
$^{1}$Swedish Institute of Space Physics (IRF), Box 812, SE-98128 Kiruna, Sweden\\
$^{2}$ Astronomical Institute of the Czech Academy of Sciences, Fri$\check{c}$ova 298, CZ-25165 Ond$\check{r}$ejov, Czech Republic\\
$^{3}$Finnish Fireball Network, Ursa Astronomical Association, Kopernikuksentie 1, Helsinki, FI-00130, Finland\\
$^{4}$Instituto de Astrofísica de Andalucía (IAA-CSIC), Glorieta de la Astronomía s/n, E-18008, Granada, Spain\\
$^{5}$Grupo de Dinâmica Orbital e Planetologia, São Paulo State University (UNESP), 12516-410, Guaratinguetá, Brazil\\
$^{6}$Valongo Observatory, Federal University of Rio de Janeiro (UFRJ), 20080-090, Rio de Janeiro, Brazil\\
}

\date{Accepted XXX. Received YYY; in original form ZZZ}

\pubyear{2025}

\begin{document}
\label{firstpage}
\pagerange{\pageref{firstpage}--\pageref{lastpage}}
\maketitle

\begin{abstract}
On 2023 February 27 at 18:15:55.77 UT, a bright fireball streaked across the sky above northern Sweden. The event offered a valuable opportunity to study the phenomenon using an optical system primarily designed for auroral studies, the Auroral Large Imaging System (ALIS\_4D), that captured the event. In this study we show the capability of ALIS\_4D to perform observations in support of meteor event analysis. We estimated the trajectory from the recorded data and computed the orbit. In addition, we investigated the origin of the meteoroid searching for its parent body. Fitting the analytical ablation model known as $\alpha-\beta$ to the trajectory as well as incorporating local wind-field data in Monte-Carlo dark-flight simulations, strewn-fields were computed and physical properties of the meteoroid were estimated.
Trajectory analyses delineate a strewn field along the border between Kiruna and Gällivare in northern Sweden. 
Our findings indicate that the meteoroid's parent body was likely an Apollo family object.
We performed an orbital similarity analysis to identify candidate parent bodies of the fireball.
Our simulations suggest that close approaches with Earth could have disrupted the meteoroid’s orbit, placing it on a collision course.


\end{abstract}




\section{Introduction}

Every day, the Earth's atmosphere is bombarded by billions of dust-sized particles and larger pieces of material from space. Objects with sizes between 100 microns and 1 meter moving in interplanetary space are called meteoroids. When a meteoroid enters the Earth's atmosphere, it ablates and produces a phenomenon called a meteor. Meteors are commonly seen as visible streaks of light in the night sky. This incoming material gives us a unique opportunity to examine the motion and population of small bodies in the solar system \citep{Vaubaillon2005A_new_method_to_predict_meteor_showers_I_Description_of_the_model, Vaubaillon2005A_new_method_to_predict_meteor_showers_II_Application_to_the_Leonids, Kastinen2017A_Monte_Carlotype_simulation_toolbox_for_Solar_System_small_body_dynamics_Application_to_the_October_Draconids}. The extraterrestrial input of matter also affects various physical and chemical processes, such as the formation of clouds at 15-25 km altitude responsible for the destruction of ozone in the polar regions, mid-latitude ice clouds at 75-85 km, which are possible tracers of global climate change, and metallic layers in the atmosphere \citep{Plane2003Atmospheric_chemistry_of_meteoric_metals, planeMesosphereMetalsChemistry2015,Silber2017}. The study of fireballs also contributes to the advancement of planetary defense. Fireballs provide a unique opportunity to examine the ability of small asteroids to penetrate into our atmosphere and the consequences of them doing so \citep{trigoRodriguezAssessmentMitigation2017,silberPhysicsMeteorGenerated2018,mainzerFuturePlanetaryDefense2021}.

To monitor and study meteors, networks such as the AllSky7 Fireball Network Europe\footnote{\url{https://allsky7.net/}}, the Global Meteor Network\footnote{\url{https://globalmeteornetwork.org/}}, the Swedish Allsky Meteor Network\footnote{\url{https://www.astro.uu.se/~meteor/}}, and the Brazilian Meteor Observation Network\footnote{\url{https://www.bramonmeteor.org/bramon/}}, among others, have been established around the world. Although these specialized networks monitor vast regions of Earth, certain meteor events are occasionally documented through unconventional means, such as surveillance cameras, dashcams, or personal recording devices. For instance, \citet{Gritsevich2014a} successfully utilized dashcam footage for meteor analysis. 

   \begin{figure}
   \centering
\includegraphics[width=.75\linewidth,trim=0cm 0cm 0cm 0cm, clip]{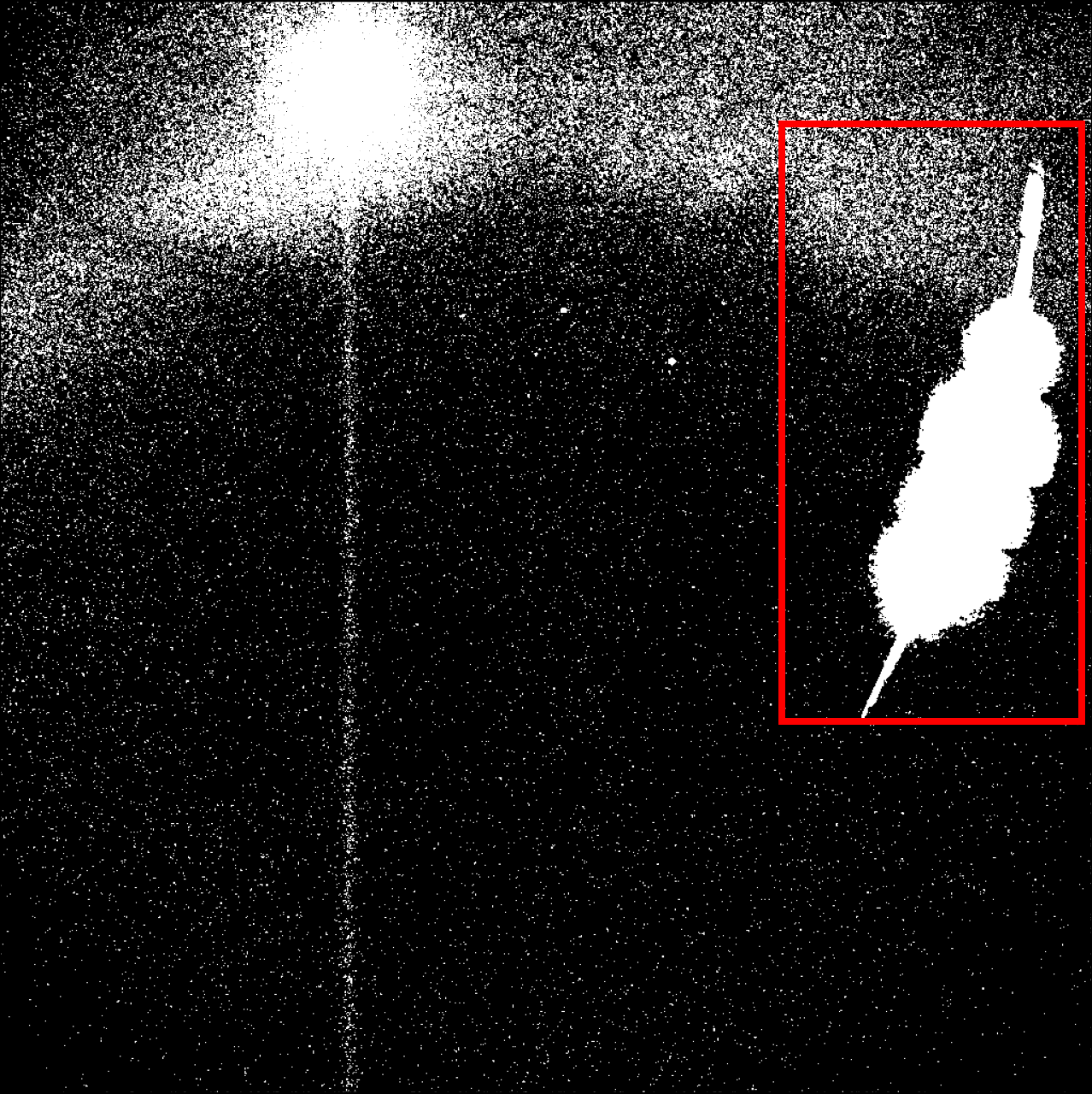}
\includegraphics[width=.75\linewidth,trim=0cm 0cm 0cm 0cm, clip]{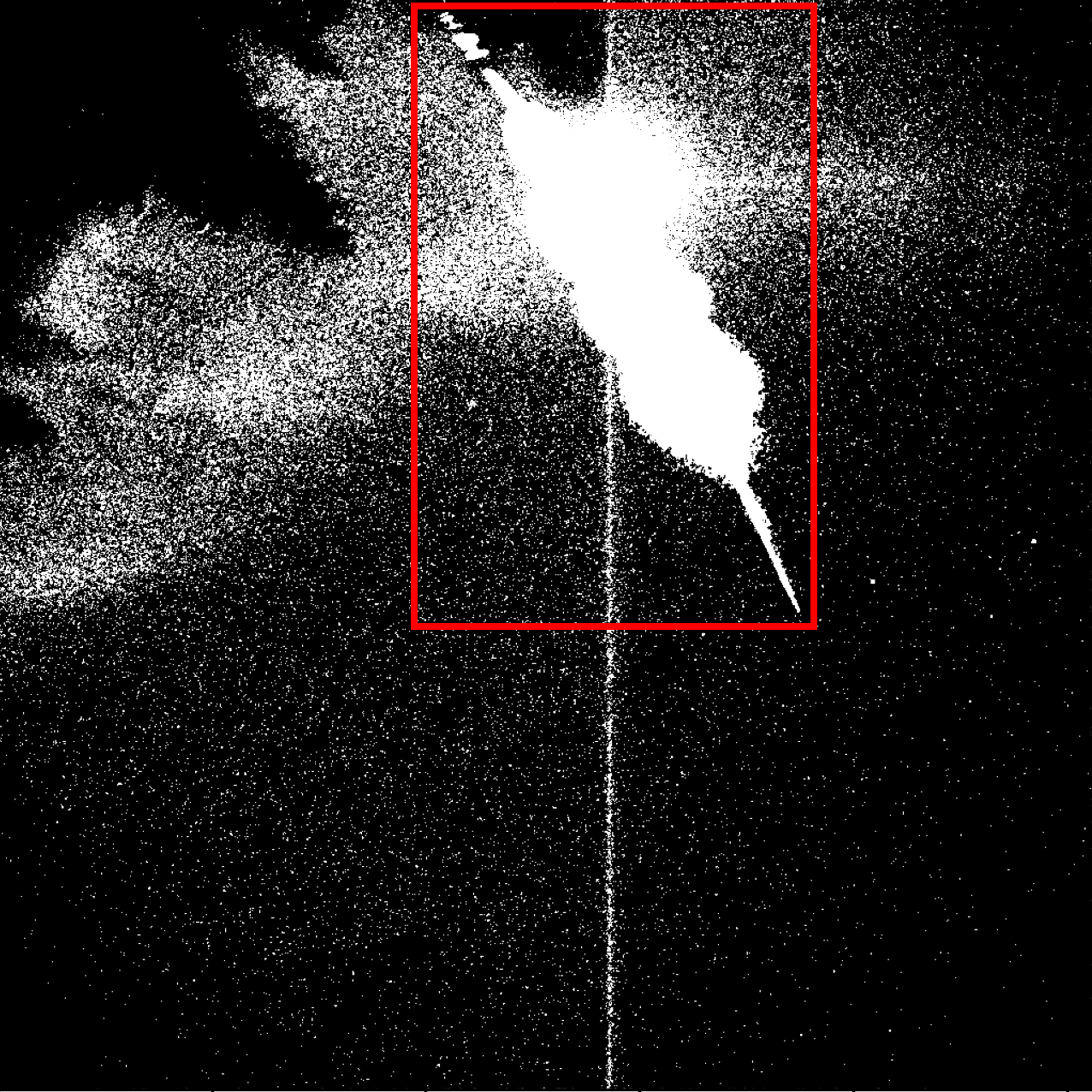}
\includegraphics[width=.75\linewidth,trim=0cm 0cm 0cm 0cm, clip]{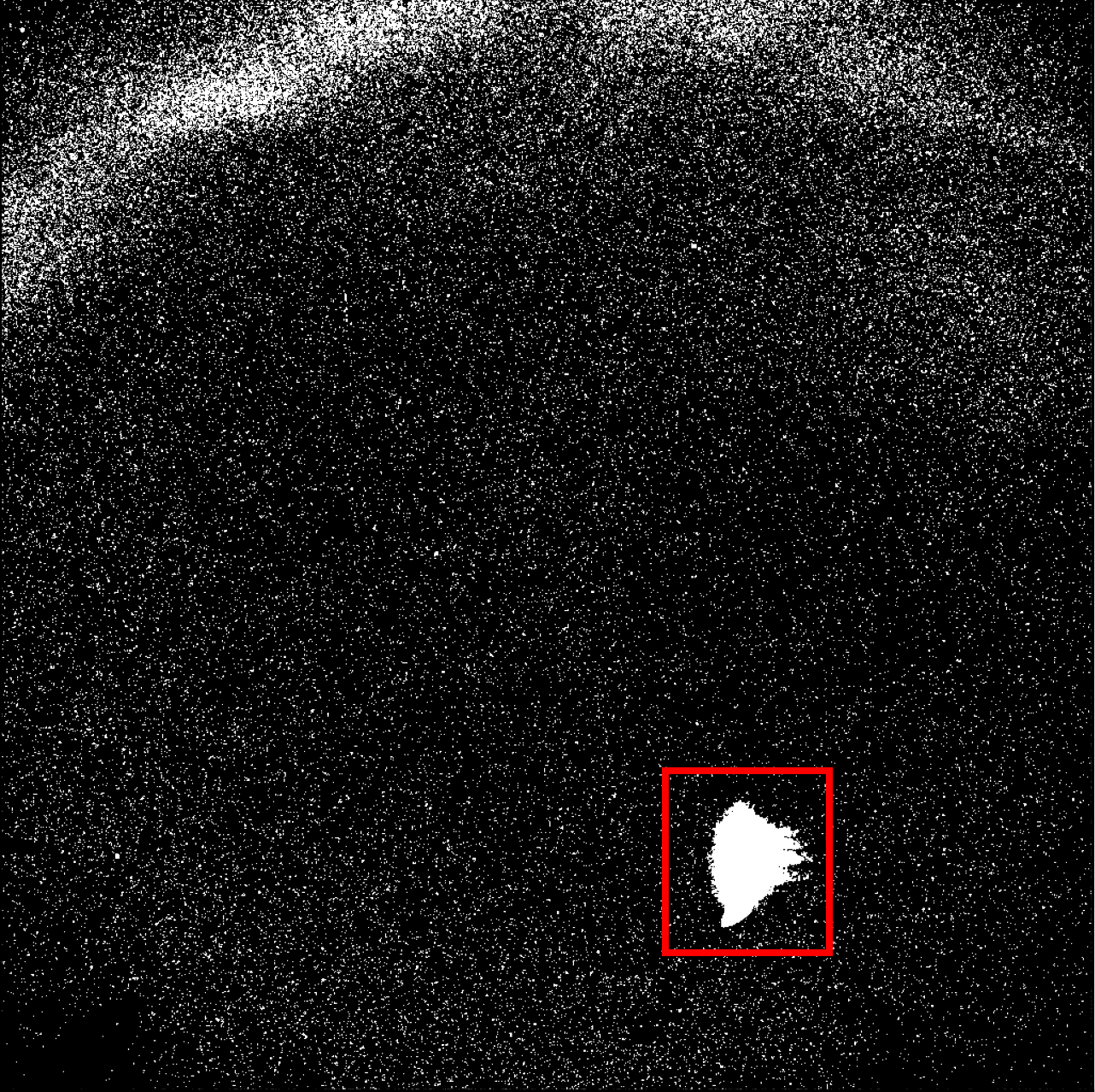}
      \caption{Composite fireball images captured by ALIS\_4D stations: Optiklab in Kiruna (top), Silkkimuotka (middle), and Tjautjas (bottom). Each image is a composite generated by stacking consecutive frames, with the fireball highlighted by a red square.}
         \label{Fig:fb}
   \end{figure}

In this work, we analyze the fireball observed on February 27 over the northern region of Sweden, primarily using images captured by an auroral optical system, an unconventional method for meteor monitoring. To present our findings systematically, we structure the paper as follows: In Section \ref{sec:obs}, we describe the conditions and characteristics under which the meteor was observed. Using the observational data, we estimate the trajectory and compute the orbit (Section \ref{Sec:AtmTrO}). Following this analysis, we predict the strewn field in Section \ref{Sec:strField}. In Section \ref{Sec:Origen}, we investigate the origin of the meteoroid by computing its similarity to minor bodies and meteor showers and by integrating the meteoroid's orbit backward in time. In Section \ref{Sec:disc}, we present a discussion of the results, and the conclusions of this work are provided in Section \ref{sec:conclusions}.

\section{Observations}
\label{sec:obs}

\begin{figure}
\centering
\includegraphics[width=\linewidth,trim=2cm 7.3cm 3cm 0cm, clip]{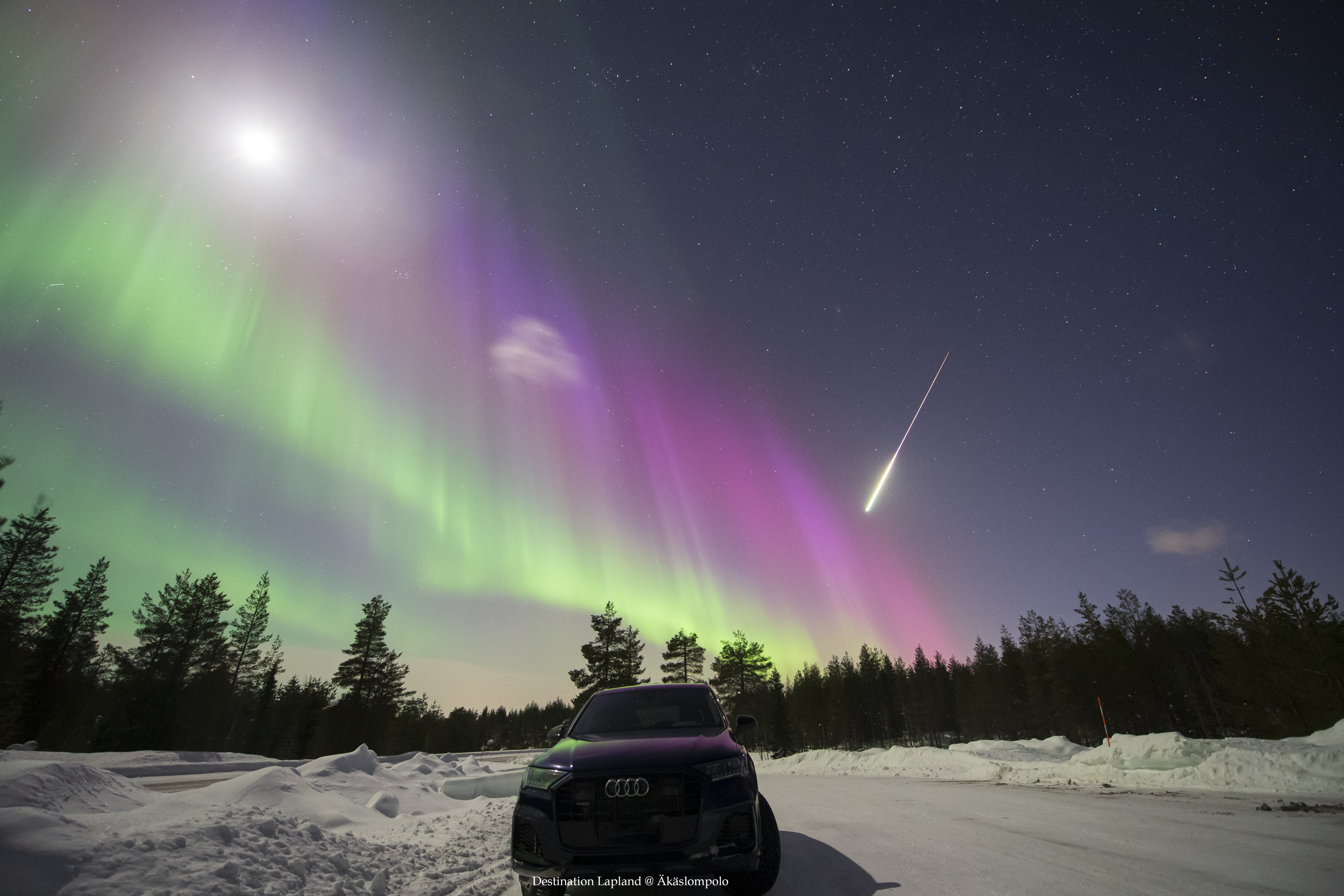}
      \caption{2023 February 27 fireball image recorded in Kittilä (credit: Mauri Kuru).}
         \label{Fig:ktl}
   \end{figure}

\begin{figure}
   \centering
\includegraphics[width=1.\linewidth,trim=0cm 0cm 0cm 0cm, clip]{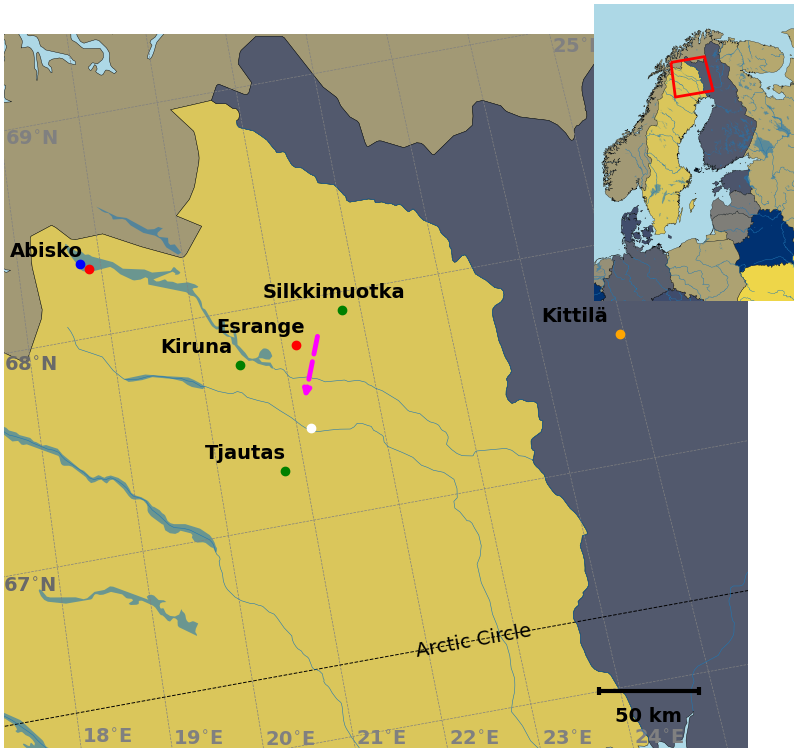}

      \caption{Map of camera positions. Green dots: ALIS\_4D stations used in this study; red dots: unused ALIS\_4D stations; blue dot: Swedish Allsky Meteor Network camera; and purple dot: privately owned camera. The magenta arrow is the projection of the fireball track on the ground, and the white dot indicates the strewn field.}

         \label{Fig:map_can}
   \end{figure}

On 2023 February 27 at 18:15 UT, a fireball appeared over northern Sweden and was detected and observed by multiple instruments (Fig.~\ref{Fig:fb}). These instruments were a Swedish Allsky Meteor Network
camera in Abisko, a privately owned camera temporarily set up along a road in Kittilä (Fig.~\ref{Fig:ktl}) and the
multi-station scientific optical system ALIS\_4D (see locations in Fig.~\ref{Fig:map_can}).

The meteor camera in Abisko is a WATEC 902H2 Ultimate with a f/2 180-degree
fisheye lens \citep{stempelsSwedishAllskyMeteor2016}. Because of the cloudiness in Abisko, the data from the Abisko camera could not be used in the further analysis, although the fireball was visible through the clouds. There is also such a
camera in Kiruna but unfortunately it was taken offline due to instrument
failure and did not record the event. The Kittilä camera was a Nikon D6,
Nikkor 14-24, Iso 10000, aperture 2.8 set to a shutter time of 2.5 seconds with
a 0.5 seconds pause between pictures. This camera was set up by Mauri Kuru, and
he describes the event as:

\begin{quote}
    "A fireball at night, as bright as the sun. I was photographing the aurora
    last night along the Ylläsjärvi Kittilä road, when suddenly a ball of fire
    appeared near the zenith, with a long rod visible. From my perspective, the
    ball disappeared to the northwest. "
\end{quote}

ALIS\_4D is a scientific infrastructure dedicated to multi-station absolute imaging measurements of narrow-band auroral emissions. It is the
upgraded version of the The Auroral Large Imaging System (ALIS) that was
operated from 1993 until 2019 by IRF \citep{Brandstrom143237}. ALIS consisted of
up to eight unmanned stations in Sweden. Each station was equipped with a
remote-controllable highly light-sensitive and high-resolution scientific CCD
detector and a filter wheel with space for six narrow-band interference filters.
This setup enabled imaging spectroscopic absolute measurements of low-light
phenomena. Although the primary target of ALIS has been aurora, it has been used
for many other applications throughout the years, including studying
differential ablation of meteors during dedicated meteor campaigns \citep{pellinen-wannberg_leonid_2004}.
Recently, an upgrade of the system was initiated, motivated by the exciting
possibilities of multi-modal observations together with the new research
infrastructure EISCAT\_3D
\citep{McCrea2015The_science_case_for_the_EISCAT_3D_radar}. The new system uses $1024\times1024$ pixels EMCCD sensors (e2v ccd201) with wider field of view optics providing a significantly higher time resolution, as compared to the old system. The imagers are equipped with a filter-wheel with several 3" narrow-band interference filters operating in visible and near-infrared, located at different stations in
northern Sweden. ALIS\_4D has several filters targeted at meteor ablation, such as
$\mathsf{Ca, Fe}$ and $\mathsf{Na}$. See Table~\ref{tab:filters} for the
complete set of available filters. Unfortunately, during the fireball event none of these filters were in use and
all ALIS\_4D stations were observing only in the "blue" auroral emission ($\mathsf{N^+_{2}}$ 1Neg\xspace) at 4278~\angstrom\xspace with a full-width half-maximum (FWHM) passband of $\pm50$ \angstrom). The ALIS\_4D stations in Abisko, Kiruna, Silkkimuotka, Esrange, and Tjautjas were active and
captured the event. However, due to cloudy skies, the data from Abisko and Esrange
could not be used in this study.

The observation of this event by the new ALIS\_4D system gives us the opportunity
to explore the feasibility of monitoring meteors using a system that was
primarily designed for a very different target: the volume emission generated by
energetic particle precipitation, i.e. aurora.

\begin{table}
    \centering
    \begin{tabular}{c|c|c} \hline
        \multicolumn{3}{c}{Differential ablation} \\ \hline
        $\lambda$ [\angstrom] & $\Delta\lambda$ [\angstrom] & Usage \\ \hline
        3950 & 92 & $\mathsf{Ca, Fe}$,Meinel \\
        4227 & 280 & $\mathsf{Ca, Fe, H_2O}$ \\
        4340.5 & 28 & $\mathsf{H_\gamma}$, Balmer \\
        4861 & 25 & $\mathsf{H_\beta}$, Balmer \\
        5893 & 200 &  $\mathsf{Na}$ \\ \hline
        \multicolumn{3}{c}{Aurora and airglow} \\ \hline
        $\lambda$ [\angstrom] & $\Delta\lambda$ [\angstrom] & Usage \\ \hline
        \textbf{4278} & \textbf{50} & $\mathsf{\textbf{N}^\textbf{+}_{\textbf{2}}}$ \textbf{1Neg}\xspace \\
        4607 & 50 & $\mathsf{Sr}$ \\
        4861 & 30 & $\mathsf{H_\alpha}$ \\
        4934 & 50 & $\mathsf{BaII}$ \\
        5577 & 40 & $\mathsf{O(^1S)}$ \\
        5525 & 40 & $\mathsf{bg}$ \\
        6300 & 40 & $\mathsf{O(^1D)}$ \\
        6562 & 25 & $\mathsf{H_\alpha}$ \\
        6700 & 25 & $\mathsf{N_2 1P}$ \\ \hline
        \multicolumn{3}{c}{Atmospheric Physics} \\ \hline
        $\lambda$ [\angstrom] & $\Delta\lambda$ [\angstrom] & Usage \\ \hline
        7320 & 23 & $\mathsf{O^+(^2P)}$ \\
        8000 & 2000 & $\mathsf{OH}$ Meinel \\
        8446 & 40 & $\mathsf{O(3p^3P)}$ \\
        8650 & 102 & \\
        12687 & 28 & $\mathsf{O2}$ IR \\
        15237 & 28 & $\mathsf{OH(1 - 3)P1(2)}$ \\
        15429 & 28 & $\mathsf{OH(3 - 1)P1(4)}$ \\
        15210 & 28 & $\mathsf{OH(3 - 1)}$ IR \\ \hline
    \end{tabular}
    \caption{Filters currently available in the ALIS\_4D system. The bold row
    highlights the 4278 \mbox{\angstrom} filter active during these observations.}
    \label{tab:filters}
\end{table}

\subsection{Data processing}

\begin{figure}
   \centering
            \includegraphics[width=\linewidth,trim=0cm 0cm 0cm 0cm, clip]{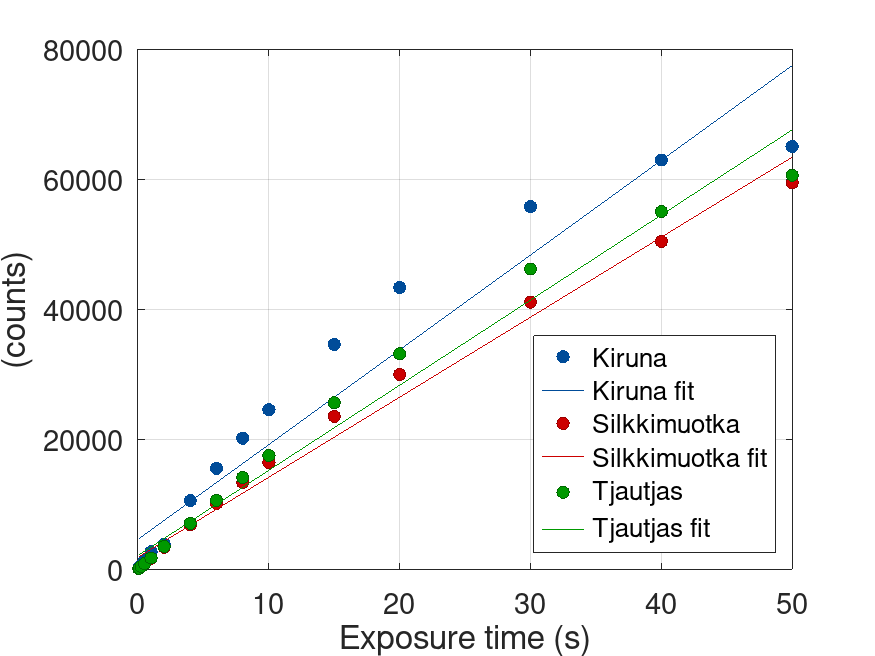}

      \caption{Linearity check using dynamic range and speed mode(Vertical Shift Speed 2.2$\mu$s; Horizontal Readout Rate 20MHz; Electron Multiplication Gain 200). Note that the values were the mean
of 64$\times$64 pixels for these measurements. The non-linear shape of these curve is similar for all four imagers. Correlation for: Kiruna is 96.7107\%, Silkkimuotka is 99.3708\%, and Tjautjas is 98.8678\%. }
         \label{Fig:linearity}
   \end{figure}

The ALIS\_4D stations stores images in the FITS-format (Flexible Image Transport System), a NASA standard in wide use in astronomy. Each FITS file contains metadata in its header detailing the capturing
conditions, while the primary data structure is a three-dimensional matrix. Two
of these dimensions correspond to the spatial resolution of the EMCCD sensor,
1024 $\times$ 1024 pixels. The third dimension represents the temporal axis, with each
slice of the matrix corresponding to a single frame acquired at uniform time
intervals. During the fireball event, ALIS\_4D operated in a mode where each
frame was captured with an exposure time of 0.098 seconds. Fig.~\ref{Fig:linearity} presents representative results from one such test conducted under a consistent operational configuration. Additional test sequences, exploring a broader range of operating modes, yielded correlation coefficients between the measured data and the fitted response curve ranging from 96.7107\% to 99.9805\%. The sensors ADC produces an unsigned 16-bit integer with a max value of 65 535, as can be seen from Fig.~\ref{Fig:linearity}, the calibration reached as close to saturation as possible without actually saturating the sensor. These results indicate that the system exhibits a high degree of linearity under the tested operating conditions, supporting the reliability of subsequent observational analyses.

The first step in our data analysis process was the extraction of objects. This is
achieved by performing frame differencing, where each frame (except the first)
is subtracted from its preceding frame in the sequence. This process reduces the
signal from static objects within the field of view (FoV), such as stars,
planets and the Moon. Subsequently, a two-dimensional convolution filter is applied
to further attenuate low-intensity signals. By this process, we improve the
visibility of objects that are moving within the FoV.

Building upon the filtered data, in the next stage we selected the frames where
the fireball is present. For each frame, we manually estimate the position of the
fireball. The provided coordinates were used to cut from the frame a
square around the fireball and by fitting the content of this square with the 2D
Gaussian function, we determined the position of the signal's centroid. The
coordinate of this centroid in the image we take as the fireball position. The
total intensity around the centroid was also integrated from the original images
after a background subtraction in order to estimate the absolute flux in the
observed spectral band.

   \begin{figure}
   \centering
            \includegraphics[width=\linewidth,trim=1.8cm 0cm 2cm 0cm, clip]{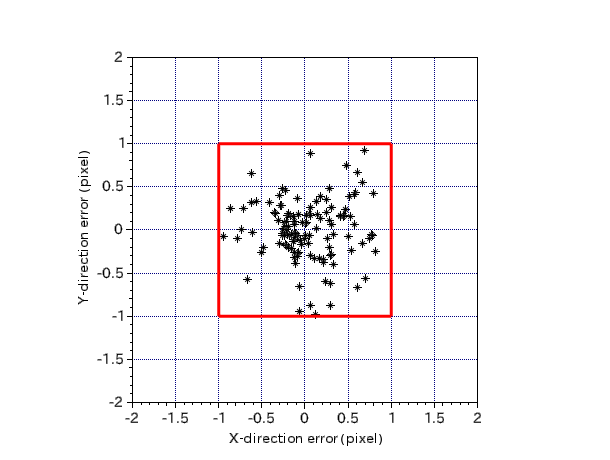}
        
      \caption{Residual of star calibration fit. The pointing error for each star on FoV is given in pixels. This calibration was performed fo the data }
         \label{Fig:pxerror}
   \end{figure}

\begin{figure}
   \centering
            \includegraphics[width=\linewidth]{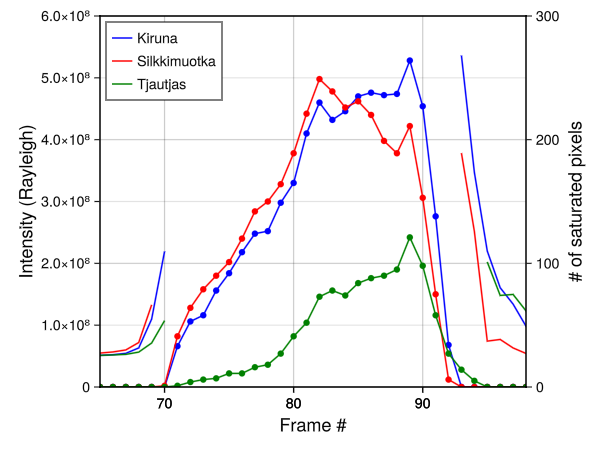}
        
      \caption{Calibrated absolute measurement of intensity in the 4278 Å (50
      Å width) filter band used at the time of the fireball event. As ALIS\_4D
      usually observes volumetric targets, the standard absolute calibration is in
      units of Rayleigh \citep{hunten1956jatp,baker1976aopt}. Only frames with no saturation are illustrated (solid lines, scale on the left-hand side) This is because the sensors saturated in the observation mode used. When there was saturation in a frame, the number of saturated pixels is shown instead (dotted lines, scale on the right-hand side). The x-axis shows the frame number. The first frame corresponds to 18:15:50 UT, and consecutive frames are separated by 0.098 s.}
         \label{Fig:flux}
   \end{figure}

The ALIS\_4D metadata includes a set of calibration parameters, such as imager orientation angles, vertical and horizontal focal lengths, vertical and horizontal displacements, and pinhole deviation. These parameters are essential for transforming pixel coordinates into other reference frames. To determine them, an astrometric calibration procedure is applied using AIDA tools\footnote{\url{https://github.com/space-physics/AIDA-tools}} developed in matlab by \citet{gustavsson2000irf} for processing ALIS image data. The AIDA tools uses the \textit{Bright Star Catalogue}\footnote{Bright Star Catalogue, 5$^{th}$ Revised Ed.: V/50 \url{ftp://cdsweb.u-strasbg.fr/pub/cats/V/50}} \citep{starcat} to perform the astrometry. For the fireball event analyzed here, the pointing residuals derived from the star calibration remain below 1 pixel throughout the field (Fig. \ref{Fig:pxerror}). Following, the analysis of the calibrated images was performed using the aidaScilab software package, a Scilab implementation of the original AIDA tools. From the fireball position and the geometrical calibration, we computed the azimuth and zenith angles of the object in the local coordinate system of each station. For these images, absolute calibration was also done, allowing the conversion of the integrated image intensity into physical units. The observed flux in the 4278 \mbox{\angstrom} band as a function of time is illustrated in Fig.~\ref{Fig:flux}. Differential chromatic refraction is not considered, given that the ALIS\_4D cameras have a lower angular resolution than the astrometric instruments. Unfortunately, the exposure time used in the experiment mode, which was running when the fireball occurred, was not designed for such bright events and the sensors were saturated almost immediately after the event started.

As this is the first analysed fireball event observed with ALIS\_4D we did not yet have a ready made pipeline for data reduction. As such, we have processed the data with the established tooling FireOwl and adapted our existing software for this purpose. The ALIS\_4D data and the image captured in Ylläsjärvi Kittilä road were processed with FireOwl, a software developed for analyzing meteor data from the Finnish Fireball Network \citep{Visuri2021,Kyrylenko2023,PeaAsensio2024,Visuri2026,Moilanen2026}. In parallel we adapted and applied \texttt{metecho}\footnote{\url{https://github.com/danielk333/metecho}}, an analysis library developed for radar meteor head echos \citep{kastinenRadarAnalysisAlgorithm2022}. The adapted code from metecho is available in a new public repository that will be used for future analysis of optical data once refactored\footnote{\url{https://github.com/danielk333/metlight}}. FireOwl results were used in subsequent processing for consistency as these were also used for the strewn-field computations. The two methods allowed for independent validation of the results between two completely different processing approaches and allowed us to construct a prototype data reduction pipeline for future observations.

\section{Atmospheric trajectory and orbit}
\label{Sec:AtmTrO}

As described above, once the trajectory detection and astrometric calibration had been done, time series of azimuths and zenith angles were available for each of the ALIS\_4D stations in Kiruna, Tjautjas, and Silkkimuotka. The long-exposure image from Kittilä (accessible via Taivaanvahti\footnote{\url{https://www.taivaanvahti.fi/observations/show/112829}}) was astrometrically calibrated using FireOwl and combined with the ALIS\_4D data to determine the characteristic parameters of the meteor and its trajectory. The methods for orbit determination and trajectory reconstruction are described in detail in \citet{Gritsevich2017,Gritsevich2024}. 

\begin{figure}
\centering
  	\includegraphics[width=\linewidth,trim=0cm 0cm 0cm 0cm, clip]{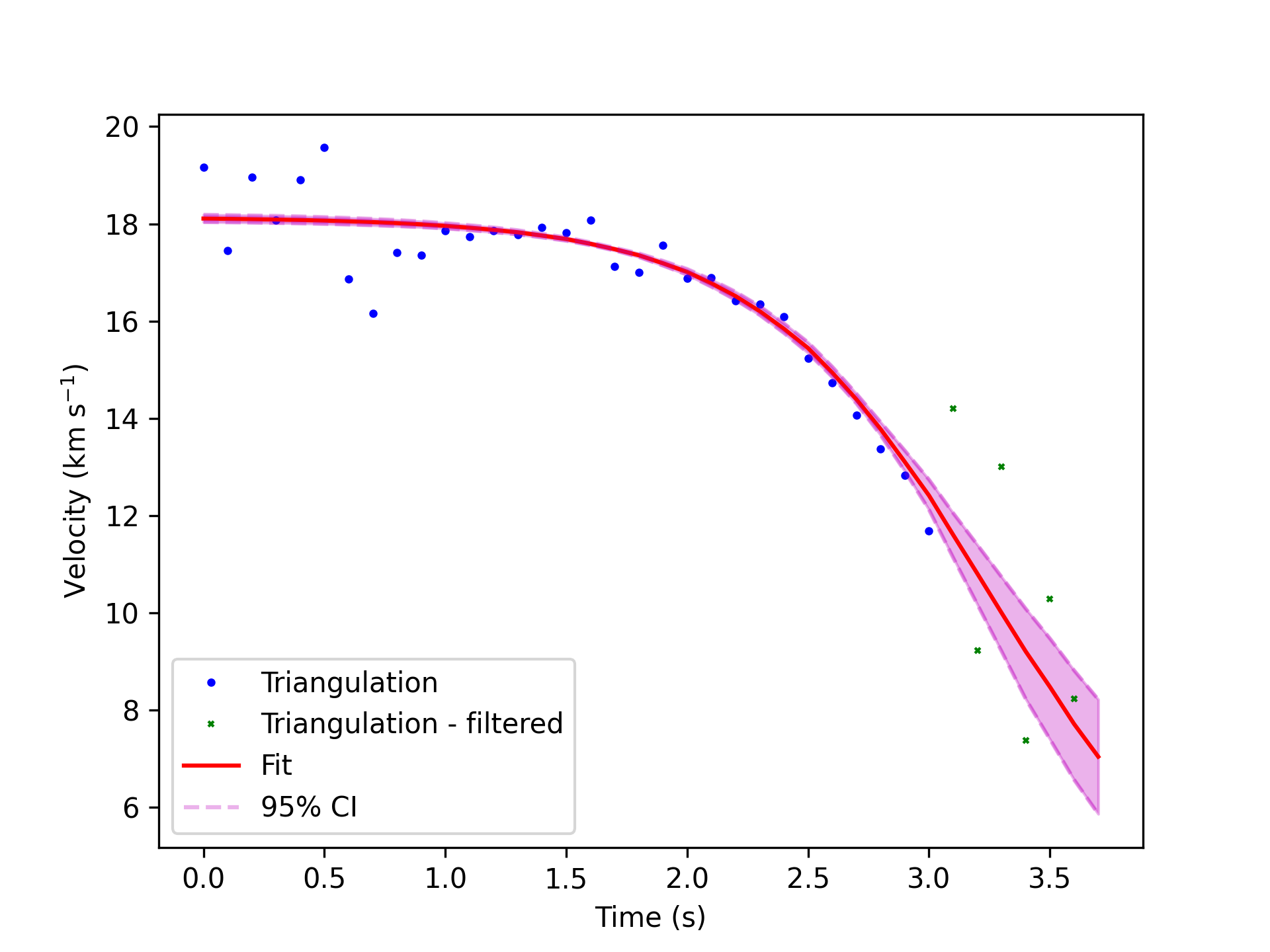}

   \caption{Velocity of the fireball during the fall. The velocities computed by the triangulation process are the blue dots on the plot. The red line is the fit obtained with metecho as described in Table~\ref{tab:trajectory} and the magenta confidence interval is obtained through direct Monte-Carlo sampling the estimated covariance matrix of the fitted parameters. The filtered data points at the end were not included in the fitting metric. Here $t = 0$ s corresponds to 18:15:56 UT.}
      \label{Fig:velocity}
\end{figure}

\begin{figure}
\centering
  	\includegraphics[width=\linewidth]{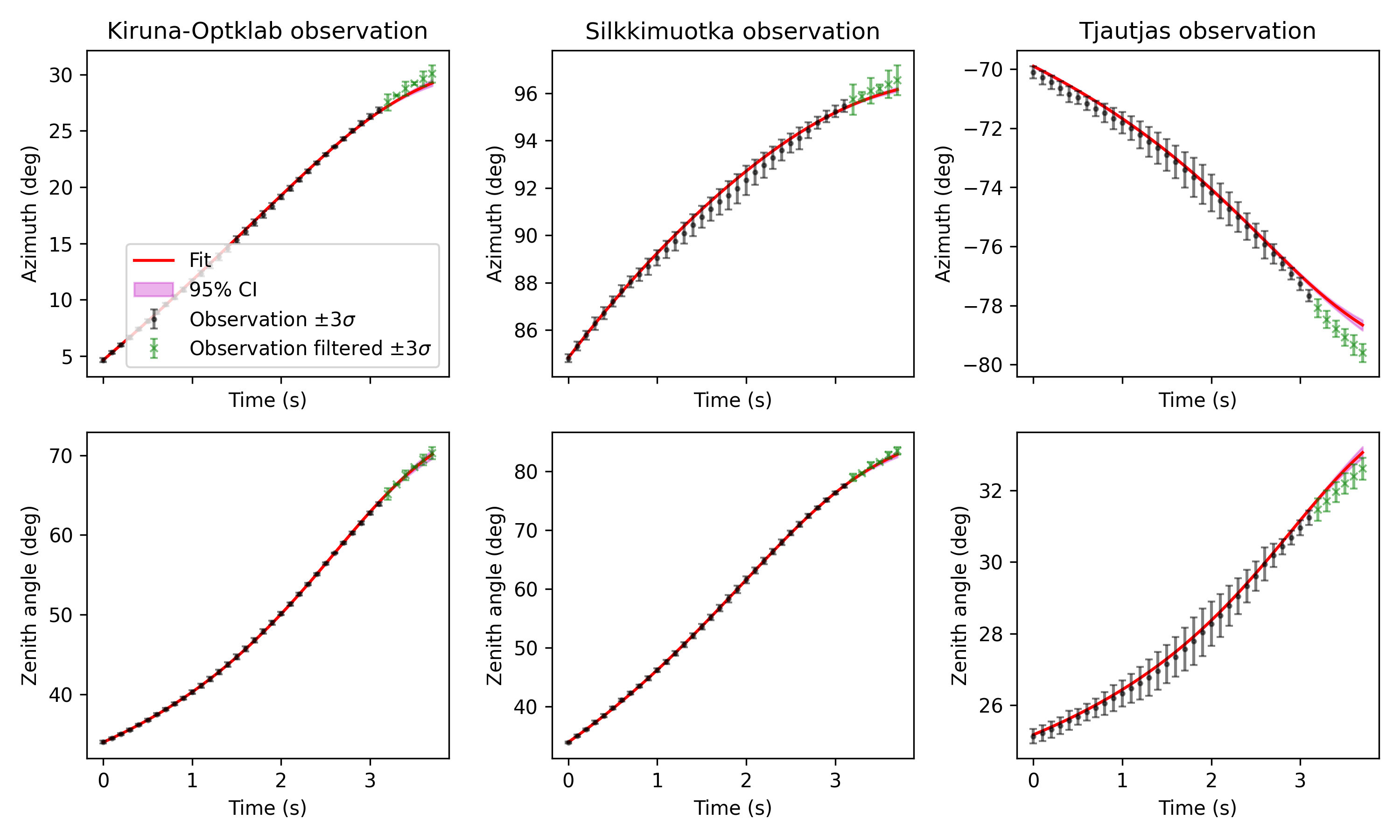}
   \caption{Azimuth and zenith angles observed by the ALIS\_4D stations together with their estimated uncertainties and the fit obtained with metecho. The uncertainties are based on the residuals of the individual frame-by-frame triangulations. The filtered data points at the end were not included in the fitting metric. Here $t = 0$ s corresponds to 18:15:56 UT.}
      \label{Fig:obs}
\end{figure}

\begin{table}
\centering
\label{tab:trajectory}
\caption[Properties]{Comparison between determined characteristic meteor parameters for the two analysis implementations.\\ $^\dagger$ The start altitude in metecho is defined by the first viable triangulation point using only ALIS\_4D and as such is lower than the one reported by FireOwl which included the Kittilä data and uses a different definition for the starting point.}
\begin{tabular}{c|ll}\hline 
Variable & FireOwl & metecho \\ \hline
$\alpha$ & 27.46 & 20.05 $\pm$ 1.43 \\ 
$\beta$ & 0.68 & 0.86 $\pm$ 0.33 \\
Entry velocity $v_e$ & 18.6~km s$^{-1}$ & 18.13 $\pm$ 0.32~km s$^{-1}$\\
Start altitude$^\dagger$ $h_0$ & 77.456~km & 72.36 $\pm$ 0.43~km \\
Slope $Y_0$ & 51.9$\degree$ & 52.4$\degree$ $\pm$ 0.6$\degree$ \\

\hline

\end{tabular}
\end{table}

The Kittilä observation yielded a high-resolution image but did not capture the
end of the luminous phase of the meteor. ALIS\_4D observations, possibly constrained by the narrow-band imaging, does not seem to have observed the end of the luminous phase
either. Given the available data the terminal height is estimated at
approximately 18 km.

Following adjustments for refractions and the implementation of delta-z
corrections following \citet{green1985spherical,Visuri2026}, the estimation of velocity and
the $\alpha-\beta$  \citep{Gritsevich2007,gritsevichPribramLostCity2008,Gritsevich2009} relies on the Optiklab-Kiruna observation only due to its advantageous orientation. Tjautjas captures the meteor from an "along"
perspective, while in the Silkkimuotka data, the fireball passed close to the
moon, which was in its first quarter phase. This might have affected the
accuracy of the centroid fitting procedure and was taken into account during
trajectory reconstruction in FireOwl. For the FireOwl reconstruction, the exponential velocity model of \citet{eqvel} (their Eq. 48), which provides a good fit to the observed data \citep{jansen-sturgeonDynamicTrajectoryFit2020a,penaAsensio2021}. Given the effectiveness of the
exponential model, the $\alpha-\beta$ model \citet{Gritsevich2009} also
demonstrates a commendable fit. The derived  ballistic coefficient  $\alpha$ and mass loss parameter $\beta$ values using the velocity measurements yield $\alpha=27.46$ and $\beta= 0.68$, as listed in Table~\ref{tab:trajectory}.

\begin{figure}
   \centering
     	\includegraphics[width=\linewidth,trim=0cm 0cm 0cm 0cm, clip]{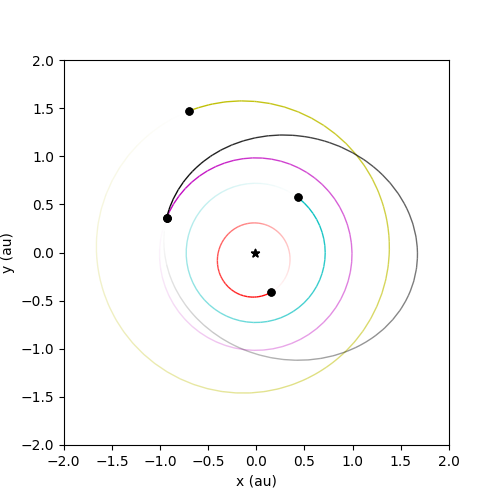}

      \caption{Last orbit of the object (black) before the collision across the inner solar system. The orbit of the planets Mercury (red), Venus (cyan), Earth (purple), and Mars (green) are represented on the plot.}
         \label{Fig:orbit}
   \end{figure}

The trajectory data of the meteor, combined with the positions of Earth and
other celestial bodies at the time of the event, provide a basis for estimating
the meteoroid's orbit. In Table~\ref{tab:orb_ele_both}, we present the computed
orbital elements, the orbit itself is illustrated in Fig.~\ref{Fig:orbit}.
This data will be used in our simulations and analyses of the meteoroid's dynamical origins. 

\begin{figure}
   \centering
     	\includegraphics[width=\linewidth]{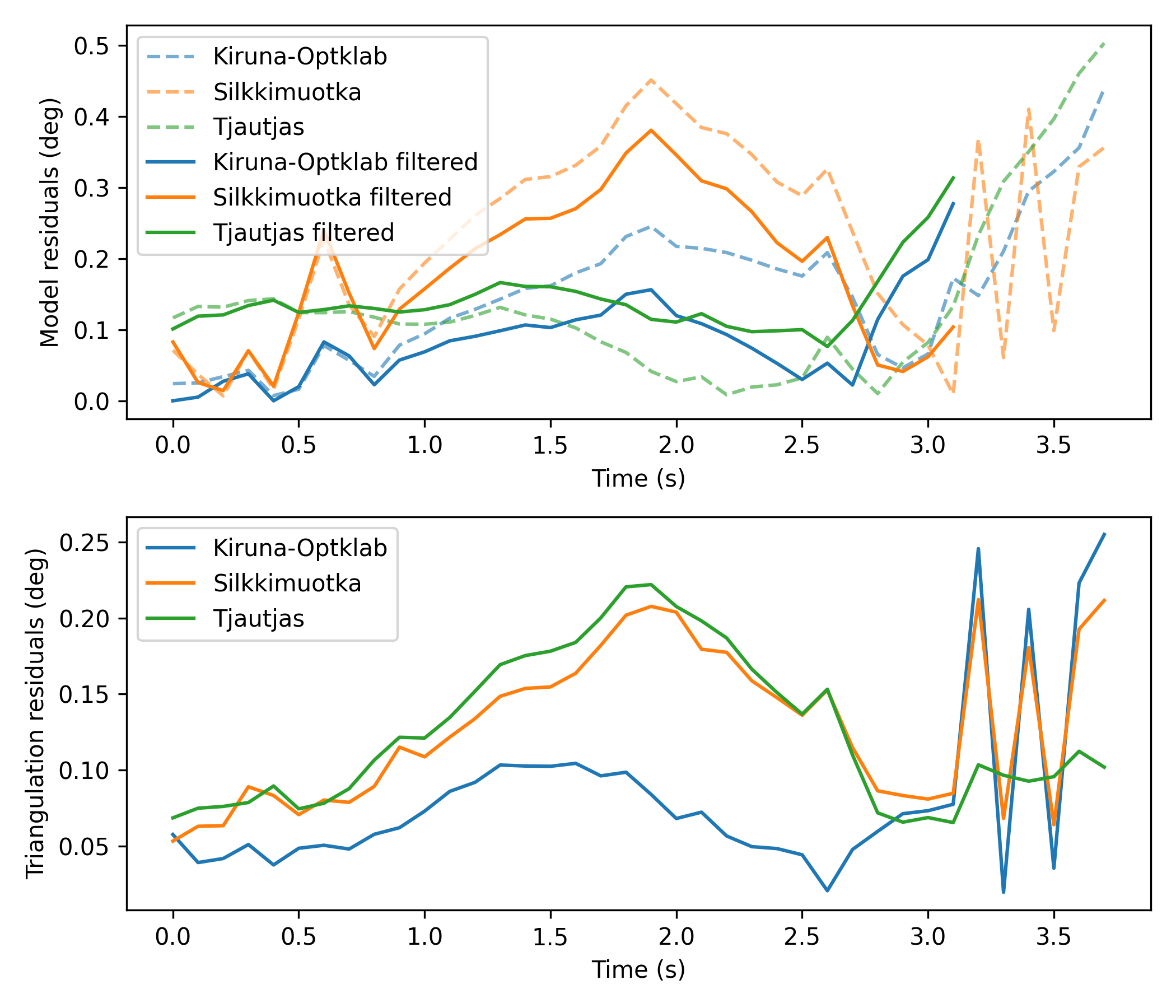}
      \caption{The upper panel illustrates the residuals between observed azimuth and zenith angles and the fitted model. The bottom panel illustrates the residuals between observed azimuth and zenith angles and the multi-station triangulation results. The residual angles were calculated as the angle between the corresponding pointing vectors.}
         \label{Fig:resids}
\end{figure}

In contrast, the metecho analysis relies on three basic components: a forward model that produces a time series of 3D positions and velocities (and other properties if avalible) for meteor, a instrument function that translates such a trajectory into observables, and the input data with estimated uncertainties. Previously, this trajectory model used ablation models or empirical formulas suitable for radar meteor sizes \citep{kastinenRadarAnalysisAlgorithm2022, Kero2008Highresolution_meteor_exploration_with_tristatic_radar_methods}. For this study we implemented a version of the $\alpha$-$\beta$ phase-space criterion that can be integrated in time allowing for a time-dependent $\alpha$-$\beta$ trajectory to be used as the forward model. Exchanging the usual radar instrument function with one that translates the trajectory to azimuths and elevations allowed fitting data from optical instruments. The ALIS\_4D time series were also used to perform frame-by-frame triangulation of the three dimensional estimated position of the fireball. These estimated positions were used to generate initial guesses for the fitting and also to estimate the empirical uncertainty in each azimuth zenith angle pair. The uncertainty of each point was defined as the angular extent between the triangulation solution and the input direction vector determined from the centroiding procedure. The fitting procedure used a regular Nelder-Mead minimization method on a least squares expression weighted by each data points uncertainty. In Fig.~\ref{Fig:resids} the residuals from the direct triangulation and the fitting residuals are illustrated.

\begin{figure}
   \centering
\includegraphics[width=\linewidth,trim=0cm 0cm 0cm 0cm, clip]{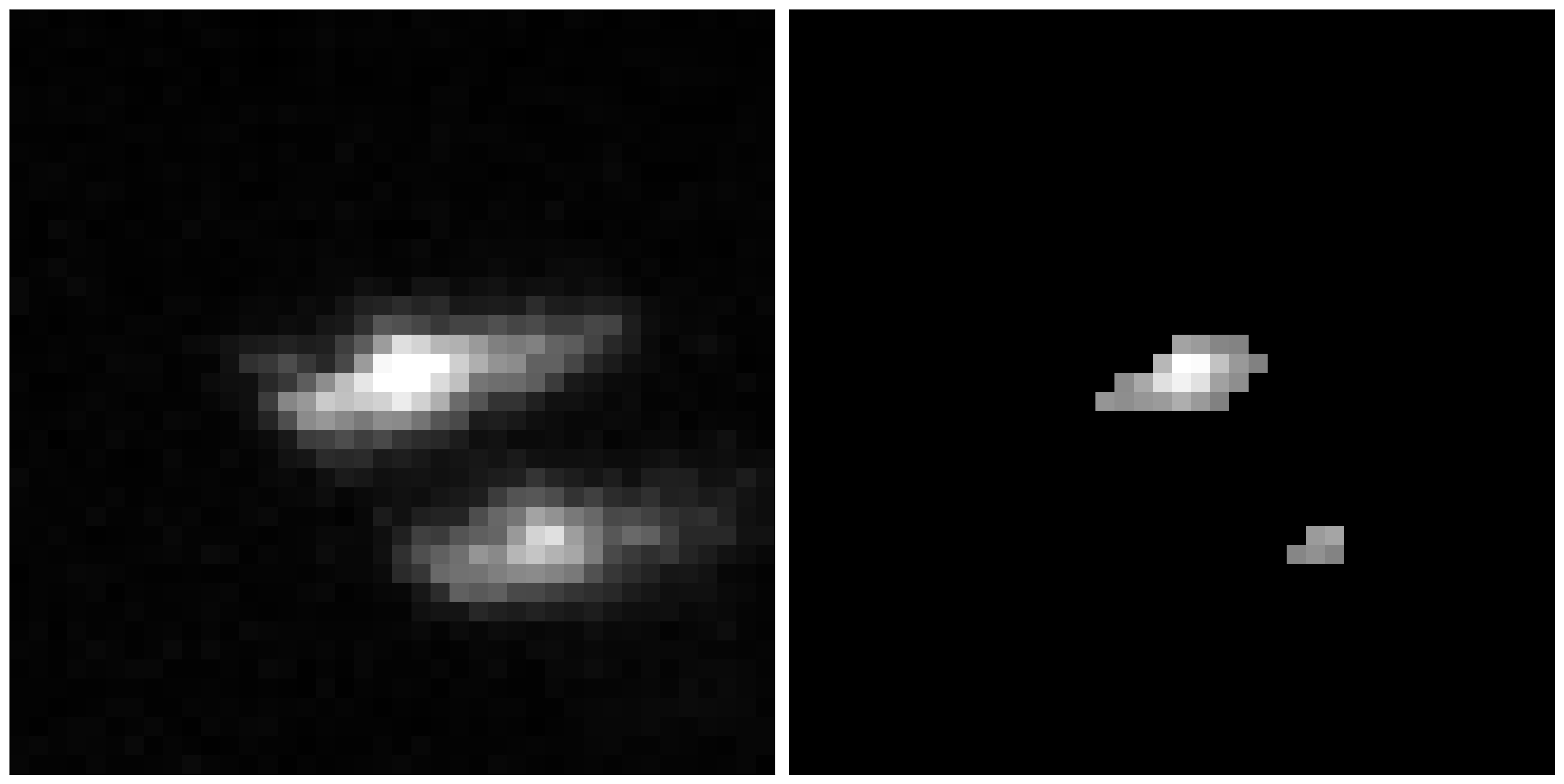}
      \caption{Zoom in on the fireball in frame 94 (time: 18:15:59.21 UT) of the image captured by the Silkkimuotka station during the fireball before (left panel) and after (right panel) post-processing for centroid detection.}
         \label{Fig:frag-frame}
   \end{figure}

In the bottom panel of Fig.~\ref{Fig:resids} a distinct change in how the error behaves can be noted at the end of the trajectory. Upon further examination we found evidence of a fragmentation event at the final stage of the fireball's luminous phase in the Silkkimuotka station's data (Fig. \ref{Fig:frag-frame}). The primary body appears to split into two distinct luminous components, which can be observed to depart from the pre-fragmentation trajectory. No clear signs of fragmentation was identified in the other stations' data, likely due to the viewing geometry constraints and the apparent separation of the fragments along their direction. As such, we excluded these points from the fitting procedure as indicated by Fig.~\ref{Fig:velocity} and Fig.~\ref{Fig:obs}. For comparison, in Fig.~\ref{Fig:resids} the residuals of a fitting without filtering is shown in the upper panel alongside the residuals from the filtered fit.

As metecho has access to both forward models for input meteor parameters and the instrument function, together with the base uncertainties, one can estimate the covariance of the estimated parameters using a linearized approach. The linearized covariance $\Sigma_{met}$ of the estimated meteor parameters is

\begin{align}
    \Sigma_{met} = \left ( J^T \Sigma_{obs}^{-1} J \right )^{-1},
\end{align}

where $\Sigma_{obs}$ is the covariance of the observations and $J$ is the Jacobian of the complete forward model. We estimate the Jacobian numerically and compute the covariance of the 8 input parameters used for the model: 6 for the start position and velocity and 2 for $\alpha$ and $\beta$. We can then sample this distribution and compute confidence intervals and distributions of derived parameters using direct Monte-Carlo. In Table~\ref{tab:trajectory} the estimated parameters together with their 1$\sigma$ interval is tabulated. In Fig.~\ref{Fig:velocity} the velocity of the fitted model, its 95\% confidence interval, and the estimated velocity using the direct triangulation, are illustrated. Similarly, in Fig.~\ref{Fig:obs}, the observations, their estimated errors, the fit and its propagated confidence interval is illustrated for the trajectory captured by ALIS\_4D.

\begin{table}
\centering
\caption[Properties]{Orbital elements of the meteoroid.The uncertainty in the longitude of the ascending node is negligible, as its value is effectively constrained by Earth’s position at the impact time for any colliding body.}
\begin{tabular}{lrr}\hline 
	
Semi-major axis (au)	& 1.344437 & $\pm$ 0.036640\\ 
Eccentricity            & 0.288285 & $\pm$ 0.018638 \\
Inclination ($\degree$) & 26.337292 &$\pm$ 0.566984\\
Longitude of ascending node ($\degree$)& 338.577982 &\\ 
Argument of pericentre ($\degree$)& 211.925412 &$\pm$ 1.260922 \\
Mean anomaly ($\degree$)& 342.709790 &$\pm$ 1.466154\\ 
\hline

\end{tabular}
\label{tab:orb_ele_both}
\end{table}

\section{Anticipated strewn field}
\label{Sec:strField}

Monte Carlo-based forward simulations of the fireball flight, followed by a dark
flight phase, were conducted to predict the possible meteorite landing area from
the 27 February 2023 fireball, following the method described in
\citet{Moilanen2021} and \citet{Gritsevich2024}. Such simulations are
particularly useful in cases where meteorites are already predicted based on
inverse analysis of the luminous trajectory
\citep{Gritsevichcite2009Classification, Gritsevich2012,Turchak2014, Sansom2019, MorenoIbez2020, Boaca2022, PeaAsensio2023, EloyMaria2025, penaAsensioConsistencyDynamicalModelingInPress},
as they provide a higher level of detail and more rigorously account for
fragmentation while avoiding unnecessary simulations for cases where the
terminal mass is zero. The model has been successfully validated through its
application to historical meteorite falls, demonstrating a strong correlation
between predicted strewn fields and the actual distribution of recovered
meteorites, both in terms of fragment masses and spatial dispersion. It has been
successfully applied to events such as the iron meteorite recovered near the
Ådalen village in Sweden, as well as the Košice, Neuschwanstein, and Winchcombe
meteorite falls, among others, yielding accurate results \citep{Moilanen2021e, moilanenJumpIronMeteorite2022, Kyrylenko2023,gritsevichStrewnFieldRibbeck2025,moilanenFirstInstrumentallyDocumented2026}.
Furthermore, the model has played a critical role in confirming several
meteorite recoveries, including Annama, Motopi Pan (asteroid 2018 LA), Ozerki, and Ischgl \citep{Gritsevich2014a,Gritsevich2014b,TrigoRodrguez2015, Lyytinen2016,Kohout2017,Maksimova2019,Jenniskens2021,Gritsevich2024}.

At the time of the 2023 February 27 fireball, atmospheric data provided by the \textit{Global Forecast System} (GFS) indicated strong winds, with speeds exceeding 24~m~s$^{-1}$ between altitudes of 40,871
meters and 673 meters. Peak wind speeds reached 56.33~m~s$^{-1}$ at 30,154 meters and
54.01~m~s$^{-1}$ at 10,061 meters altitude. These strong winds reduce the accuracy of
the atmospheric model and cause significant variation in wind conditions
experienced by different fragments, resulting in a drifted strewn field.

The DFMC simulation began at an altitude of 41.09 km, during the luminous phase
of the flight trajectory (Table~\ref{tab:DFCM}). Shortly after initiation, a
fragmentation event broke the meteoroid into smaller fragments. The deceleration
value ($a_0$) at this stage was used to estimate the maximum possible meteoroid
mass at that point, setting an upper limit for the total mass of simulated
fragments. Each fragment was assigned an individual wind profile and tracked
independently to the ground \citep{Moilanen2021e}. Due to ablation, not all
fragments survived the descent.

\begin{table}
\caption[DFMC]{The DFMC input parameters for the 2023 February 27 18:15:55~UT fireball.}
\begin{tabular}{llll}\hline 
Parameter                  & Mean value     & Error           & Remarks \\\hline	
longitude ($\lambda _{0}$) & 21.1532$\degree$E & (see $e_0$)     & WGS84. \\ 
latitude ($\varphi _{0}$)  & 67.7089$\degree$N & (see $e_0$)     & WGS84. \\ 
altitude ($h _{0}$)        & 41.09 ($km$)    & (see $e_0$)     & - \\ 
spatial                    & 0 ($m$)        & $\pm$300 ($m$)  & Spatial error of the  \\
error ($e _{0}$)           &                &                 &  start point.\\ 
trajectory                 & 206.4$\degree$    & $\pm$1.8 $\degree$ & 0$\degree$-360$\degree$ \\
direction ($\delta _{0} $) &                &                 &(0° = N, clockwise). \\ 
trajectory                 & 51.864$\degree$   & $\pm$1.8 $\degree$ & 0$\degree$-90$\degree$ \\
slope ($Y _{0}$)           &                &                 &(90$\degree$= vertical).\\ 
velocity ($V _{DF0}$)$^{\textbf{a}}$        & 16 562.3       & $\pm$300        & Dark flight initial \\
                           & ($m$ $s^{-1}$) & ($m$ $s^{-1}$)  &velocity. \\
deceleration               & 2 804.254      & -               & Deceleration at the  \\
limit ($a _{0}$)           & ($m$ $s^{-1}$) &                 & start point.\\ 
fireball end               & 18.00 ($km$)   & -               & Ultimate cut-off value \\
height ($h _{t}$)$^{\textbf{b}}$ &                &                 & inferred from our\\
                           &                &                 & analysis. \\ 
bulk                       & 3.3            &$\pm$0.5         & Default  values for\\
density ($\rho _{m} $)     & ($g$ $cm^{-3}$)&($g$ $cm^{-3}$)  & a chondrite. \\ 
ablation                   & 0.0018         & -               & After \\
coefficient ($\sigma $)    & ($s^{2}$ $km^{-2}$)&             &\citep{Moilanen2021} .\\ 
ground                     & 0.0 ($km$)      & -               & Ground elevation\\
level ($g _{e}$)           &                &                 &in simulation.\\ \hline

\end{tabular}
$^{\textbf{a}}$\hspace{0.5em}\parbox[t]{\linewidth}{$V_{\mathrm{DF0}}$ is the initial dark-flight velocity used in the DFMC simulation, derived from luminous-phase data from the Kiruna Optiklab station. The selected value corresponds to a point near the end of the trajectory, but not the terminal stage, to avoid uncertainties from progressive fading while preserving dynamical consistency with the onset of dark flight.}

$^{\textbf{b}}$\hspace{0.5em}\parbox[t]{\linewidth}{End-of-visible-flight altitude: 18,390 m (plane-intersection method). The DFMC terminal height $h_t$ is set slightly below the observed value to ensure ablation \citep{Moilanen2021}.}
\label{tab:DFCM}
\end{table}

The DFMC simulation results indicate a nominal initial mass of 1,234 grams from
the 41.09 km starting point. After ablation, 969 grams (78.5 per cent) reached the
ground. This nominal mass is a reference value that does not undergo
fragmentation and follows the atmospheric conditions without Monte Carlo
variations (see \citet{Moilanen2021} for additional detail). Individual
simulation runs indicate an average of 8 to 9 fragments reaching the ground,
with sizes ranging from 3 to 627 grams.

\begin{figure}
   \centering
     	\includegraphics[width=\linewidth,trim=0cm 0cm 0cm 0cm, clip]{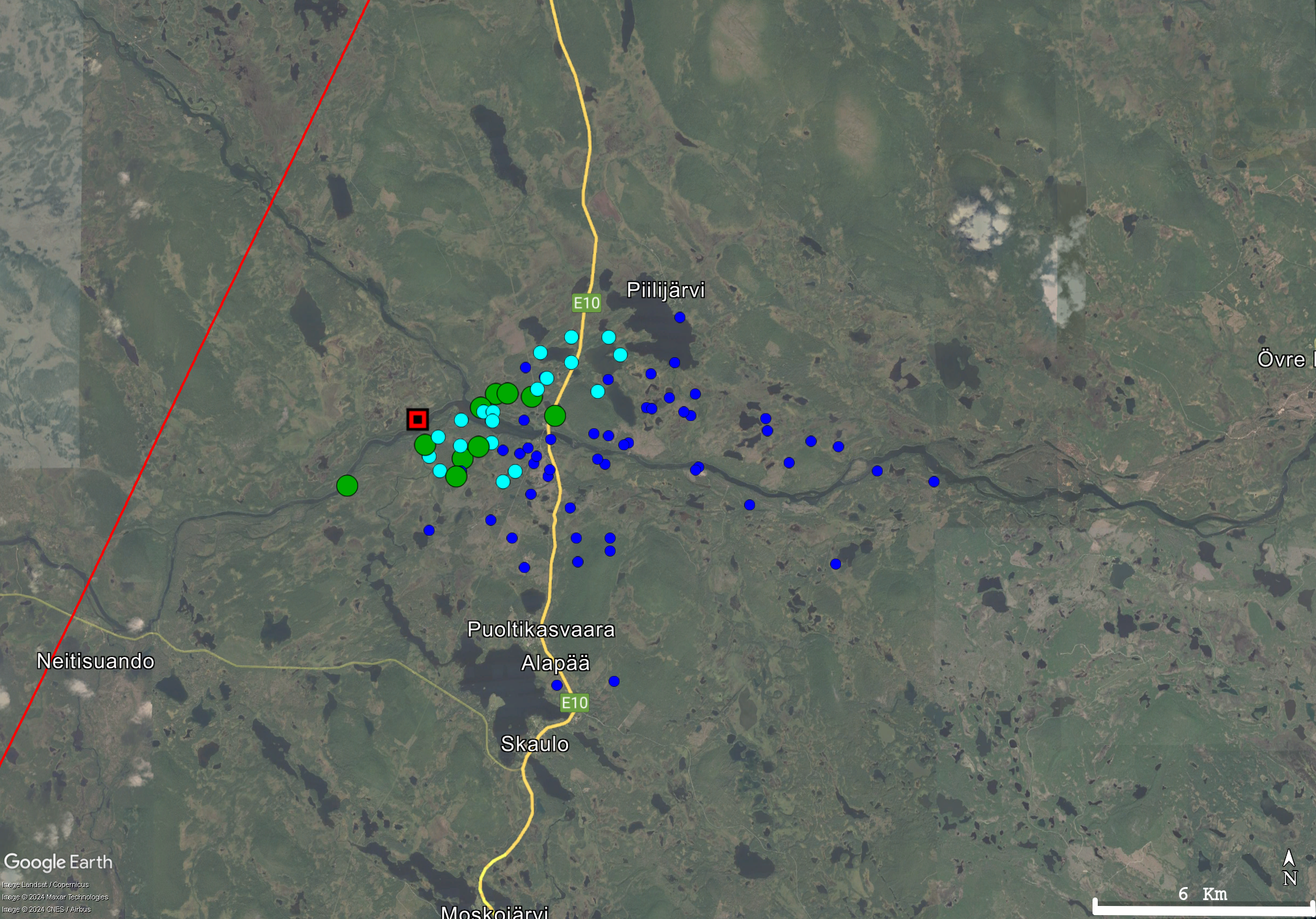}

      \caption{A DFMC simulation following the model of \citet{Moilanen2021} was conducted using the parameters listed in Table~\ref{tab:DFCM} The red box (N) marks, for reference, the impact site for the virtual nominal mass that remained unfragmented. Simulated fragments are color-coded as follows: green for 0.3–1 kg, cyan for 0.1–0.3 kg, and blue for fragments smaller than 0.1 kg. The red line on the left represents the initial trajectory of the observed fireball in the simulation.}
         \label{Fig:strewnfield}
   \end{figure}

To create a comprehensive map of the strewn field, ten individual DFMC
simulation runs were combined in Fig.~\ref{Fig:strewnfield}. This "burn"
simulation generated a higher number of fragments than a single simulation
would, allowing for a more precise definition of the strewn field's full extent.
Due to inherent uncertainties in atmospheric data, the predicted area may not be
entirely accurate. However, it is highly likely that at least some meteorite
fragments have landed within the designated region. This level of accuracy is
sufficient for the primary goal of meteorite recovery teams: locating the first
fragments, as their discovery confirms the strewn field’s position and
facilitates further recoveries.

\section{On the Origin of the fireball}
\label{Sec:Origen}

After determining the orbit that led the object to fall on Earth, we can address
the question of the meteoroid's origin. To investigate this, we employed a set of
analytical tools and methodologies, such as the method of Drummond \citep{Drummond1981} and the method of 
Nesvorn\'y \citep{Nesvorn2006} methods.

The method proposed by Drummond focuses on bodies associated with meteor showers
using a so called d-criterion \citep{Drummond1981}. The d-criterion measures the
total difference between two orbits through the variations in their orbital
elements. As a reference, a ``d'' value below 0.10 indicates similarity
\citep{Drummond1981}. On the other hand, the Nesvorn\'y method aims to identify
asteroid families, particularly those that formed recently\citep{Nesvorn2006}.
The approach relies on osculating orbital elements, which are sensitive to
short-term variations and are, therefore, suitable for identifying young
families. It applies the Hierarchical Clustering Method (HCM), which identifies
compact groups of asteroids in a multidimensional space based on their orbital
elements. The method also involves analysis of the convergence on orbital
angles, which may indicate a fragmentation event of the parental body. In
addition, Nesvorn\'y accounts for the Yarkovsky effect. Both approaches are complementary, as they probe different aspects of the orbital configuration: the Drummond criterion emphasizes angular elements, while the Nesvorn\'y method accounts for orbital energy. Their combined use therefore strengthens the assessment of orbital similarity.

To investigate whether the object is associated with a known meteor shower, we used Meteor Integrator Program (PIM)\footnote{\url{https://github.com/gdop-unesp/PIM.git}}.
The derived radiant, obtained from the projection of the atmospheric trajectory onto the celestial sphere, is located at right ascension 160.40$\degree$ and declination +8.29$\degree$ (J2000.0), with an uncertainty of $\pm 0.37\degree$. Table~\ref{tab:sim_met_show} lists the angular distances to the three nearest meteor-shower radiants. Subsequently, we calculated the Drummond criterion (d-Drummond) for each of these three meteor showers. The d-Drummond values suggest that the fireball is
unlikely to be associated with any of the considered meteor showers.

\begin{table}
\caption[Properties]{List of the three smallest angular distances between the fireball radiant and meteor showers radiant ($\delta \theta$). }
\begin{tabular}{lr}\hline 
Meteor Shower	&$\delta \theta$ \\ 

\hline	
nu-Hydrids             & 1.60023$\degree$ \\
rho-Leonids            & 5.67576$\degree$ \\
April beta-Sextantids  & 5.92947$\degree$ \\  \hline
\end{tabular}
\label{tab:sim_met_show}
\end{table}

In the absence of a strong meteor shower match, we extended our analysis to
evaluate potential orbital similarity with known minor planets. For this we used
the pipeline described in Appendix~\ref{app:a}. Table~\ref{tab:sim_ast} presents
the five objects with the highest similarity with the fireball considering
Drummond, Nesvorn\'y and both combined.

\begin{table}
\caption[Properties]{Similarity with asteroid using Drummond and Nesvorn\'y d criteriums. List of the five parental body candidate with lower value of d-Drummond and the five with the lower d-Nesvorn\'y criterion, and the list of the parental body candidates with lower values in both d-Drummond and d-Nesvorn\'y(Overlapping Candidates).  }
\begin{tabular}{lrr}\hline 
\multicolumn{3}{c}{Drummond}
\\ \hline
Minor Planet	&d-Drummond&\\ \hline	
2010 CR19&	0.041396 &\\
2019 RN1&	0.041921 &\\
2014 DH6 &	0.045822&\\
2020 DY3&	0.049541&\\
2017 NQ6&	0.056409&\\ \hline

\multicolumn{3}{c}{Nesvorn\'y}
\\ \hline
Minor Planet	&d-Nesvorn\'y (m~$s^{-1}$)&\\ \hline	
2010 CR19 &	698.662482&\\
2014 DH6 & 1068.627795&\\ 
2007 FC3 &	1151.900789&\\ 
2017 YC8 &	1292.255705&\\
2019 TP1 &	1640.774791&
\\ \hline
\multicolumn{3}{c}{Overlapping Candidates}
\\ \hline
Minor Planet & d-Drummond &d-Nesvorn\'y\\ \hline	
2010 CR19 & 0.041396 & 698.662482 \\ 
2014 DH6 &  0.045822 & 1068.627795 \\ 
2017 NQ6 &  0.056409 & 1699.391925 \\
2017 YC8 &  0.063734 & 1292.255705\\ 
2007 FC3 &  0.077933 & 1151.900789 \\ 
 \hline

\end{tabular}
\label{tab:sim_ast}
\end{table}

The results presented in Table~\ref{tab:sim_ast} suggest that 2010 CR19 is most likely the parental body of the investigated fireball. According to both the Drummond and Nesvorn\'y criteria; both
objects are classified as Apollo-type objects.

Another approach to investigate the origin of the fireball is to analyze the
possible orbital dynamics of such a body. To do this, we integrate 500 years
backward 3000 clones of the original body. The clones are test particles with
initial conditions close to the fireball ones. The initial conditions are
randomly distributed in a range of 0.028 km around the fireball semi-axis,
around 0.014 in its eccentricity, and around 0.003 radians in its inclination.
These ranges were chosen based on the orbital element errors computed (Tab.
\ref{tab:orb_ele_both}).

\begin{figure}
   \centering
     	\includegraphics[width=\linewidth,trim=0cm 0cm 0cm 0cm, clip]{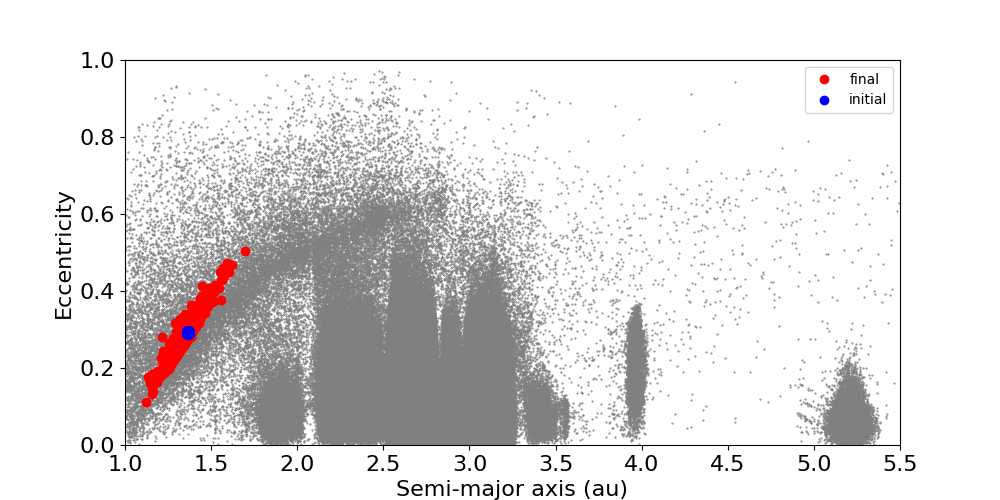}
      \includegraphics[width=\linewidth,trim=0cm 0cm 0cm 0cm, clip]{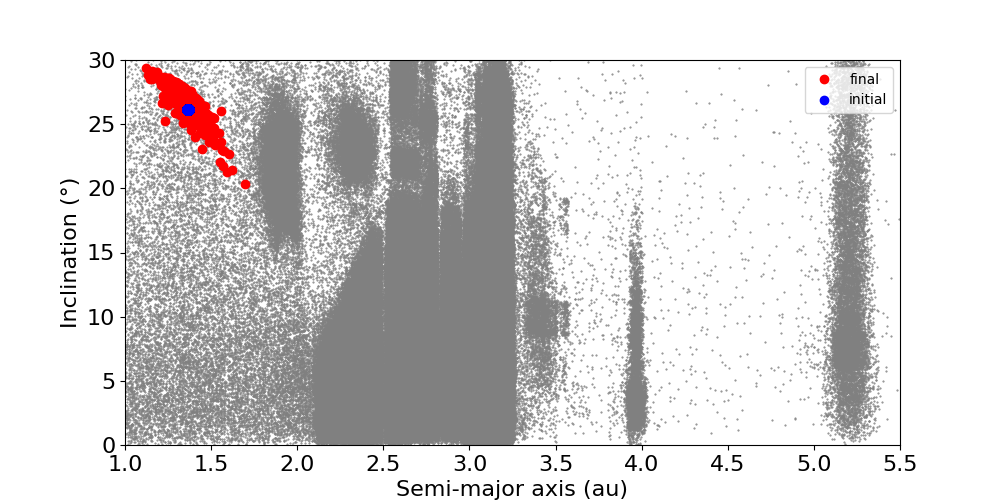}

      \caption{Locations of the remaining clones from the cloud in relation to other objects in the solar system(in gray). The blue dots are the clones at the beginning of the simulation and the red dots are the clone position at the end, i.e., 500 years before.
}
         \label{Fig:e-ixa}
   \end{figure}

We compared the orbits of the clones with the orbits of cataloged small objects
in Fig.~\ref{Fig:e-ixa}. The plot of the position of the clones, 500 years
before the collision, match with the NEA population. The clones are dispersed
throughout all asteroid family regions.

\begin{figure}
   \centering
     	\includegraphics[width=\linewidth,trim=0cm 0cm 0cm 0cm, clip]{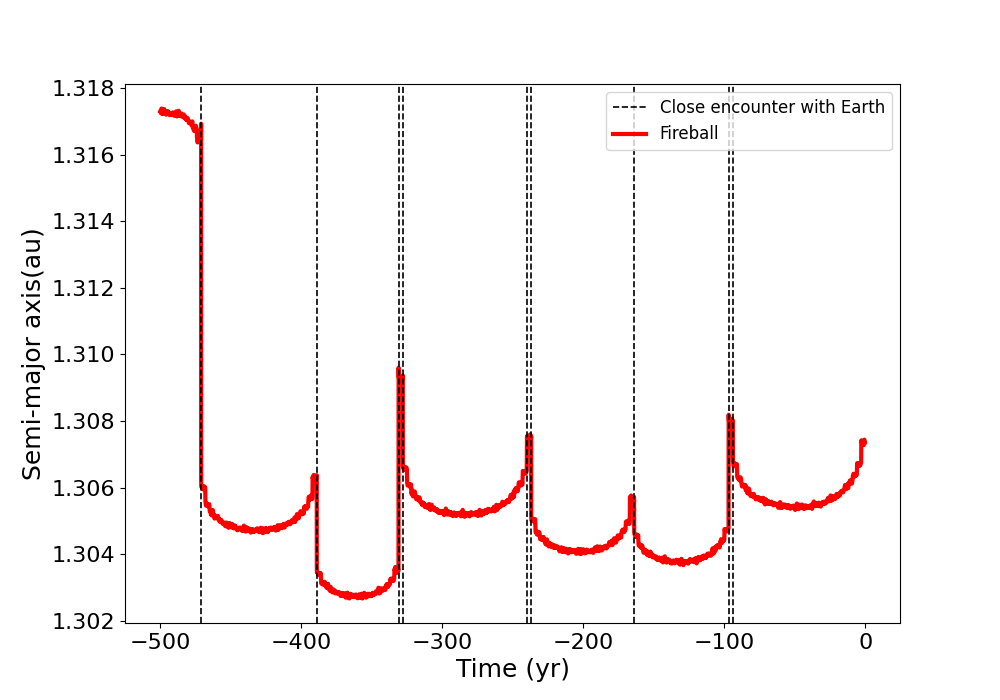}
      \includegraphics[width=\linewidth,trim=0cm 0cm 0cm 0cm, clip]{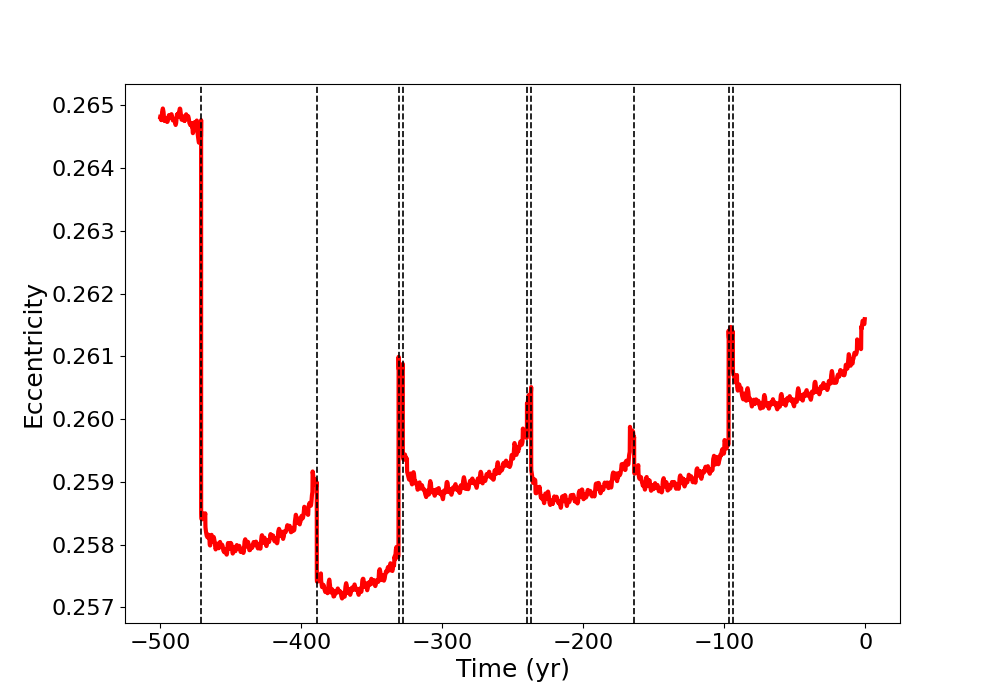}
          \includegraphics[width=\linewidth,trim=0cm 0cm 0cm 0cm, clip]{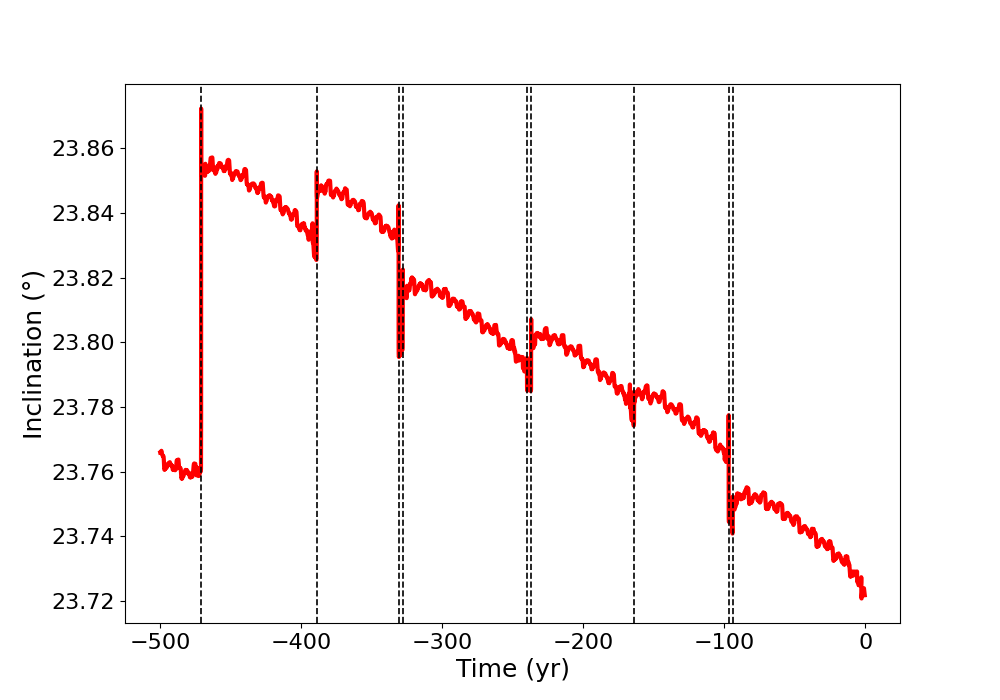}
      \caption{Evolution of orbital elements of the fireball body from 500 years before until the collision. Top panel shows the evolution of the semi-major axis, the middle panel show the eccentricity, and the bottom panel the inclination. The dashed lines marks close encounters of the object with Earth. }
         \label{Fig:aei500}
   \end{figure}

Fig.~\ref{Fig:aei500} shows the evolution of the semi-major axis,
eccentricity, and inclination during the 500 years. During the simulation, the object had 9 close encounters with Earth. These approaches disturbed the orbit
and may have put the object on a collision course with Earth.

\section{Discussion}
\label{Sec:disc}

This event highlighted the advantage of having dedicated
meteor cameras around advanced scientific instrumentation such as ALIS\_4D. As
mentioned, there was also a meteor camera in Kiruna but it was not operational
at the time due to instrument failure. Having both the wide-band visible light
observations together with the narrow-band data would have allowed for correlations between
parameters such as terminal height and a complete light curve during saturation. As the meteor cameras have a standardized
event detection already implemented, they can also be used to mine data from other
instruments in the region.

The general rule of thumb formula presented in
\citep{gritsevichStandardsCraterFormation2011,Sansom2019} for classifying fireballs as
potential meteorite droppers indicates that this event could have produced
meteorites. Following this, the strewn field calculations confirmed that
meteorites were highly possible. Unfortunately no standardized pipeline
for such events were in place at the time of the fireball. As such, several
months elapsed between the actual event and the determination of the strewn
field. A few search attempts were made but no fragments were recovered during
those excursions. Although no fragments were recovered, the event crystallized
the analysis process which should allow for swift response in the future. 

Additionally, the successful analysis of a fireball event with ALIS\_4D opens up
for targeted campaigns using the specific meteor filters with this new upgraded
system. The versatile experiment description system in place at ALIS\_4D allows arbitrary sequences of filters to be cycled at each individual station. Although the observations described here saturated the sensor, the system is capable of capturing at shorter exposure, lower electron multiplication (EM) gain, and by using different narrow-band filters, which could allow for the complete reconstruction of the intensity curve of bright events. It is also possible to reconstruct the region of light emission around and behind the meteoroid using tomography, given that the region is large enough to be resolved. As the filters are available in a filter wheel, one could for example cycle between measuring different emission ratios by employing different filters at different stations simultaneously. Such observations can provide a unique dataset for examining ablation processes. Although we can extract absolute photon fluxes in
terms of Rayleigh, it remains to process this information in combination with
other measurements to yield useful information about the ablation process. As long as absolute calibration measurements, such as the ones shown in Fig.~\ref{Fig:linearity}, are available, observed counts can be converted to absolute fluxes, even close to saturation. However, at saturation it is still difficult to extract useful information. As such experiments should be designed to avoid this issue. While some nonlinearities are present in the calibration, these can be compensated for using the same data by fitting higher order functions to the results. In this study, as the flux observations were not used in subsequent analysis, this was considered future work.

Northern Scandinavia currently host a vast amount of scientific instruments due
to its unique geographic location and due to the presence of vital research
infrastructure such as the EISCAT radars
\citep{Vierinen2022,
pellinen-wannbergForthcomingEISCAT3D2016,eiscat2024}.
Simultaneous observation with these other instruments, such as meteor camera
networks, meteor radars, spectral imagers, etc, enable novel multi-modal
studies. For example, once the new research infrastructure EISCAT\_3D is online,
simultaneous optical and multi-static radar measurements observations can be
used for detailed studies on meteor ablation and the produced meteoric plasma.
In addition, as ALIS\_4D provides absolute observations of the photon flux,
combined observations with a dedicated meteor wide band spectral imager, such as
the one presented in \citet{borovickaSurveyMeteorSpectra2005}, could be
considered for deployment close to the common observation volume. 

The data produced by ALIS\_4D in combination with other optical system was enough for the successful computation of the fireball's trajectory and thereafter the meteoroid' s orbit. Nonetheless, both can be equally computed using only data from ALIS\_4D. Once the stations operate synchronously, capturing the event with at least two stations is sufficient to determine the object's position at each recorded moment. The wide FoV introduces an error that increases with the distance between the object and the station, which could impose a limitation on the use of ALIS\_4D. However, this constraint can be mitigated by incorporating additional stations and fitting the light signal (as done in this work) to overcome the pixel resolution limitation.

It is important to note that the linearized covariance estimate only describes the covariance of the fitted parameters given the distribution of data and the used models. It says nothing about the correspondence between those models and reality. Something that can illuminate such correspondence are the fitting residuals. In this case, the fit itself is quite stable but the residuals indicate that we are not limited by the quality of the data but rather the choice of models, i.e small confidence intervals on observables compared to the residuals. These discrepancies are probably due to the centroiding during saturation and the presence of significant fragmentation. As future work, one should aim at maximizing the complexity of the applied trajectory model while still ensuring its fitting is supported by the data.

\citet{Shober2024} highlighted the challenge of assessing similarity between meteors and NEOs, due to the variability in the NEOs' orbital elements arising from their chaotic dynamical behavior. This is our case, since the fireball was a NEO in the region of Apollo family. The shorter the time between the ejection and the fireball fall, the less the similarity between the fireball and its parent body varies. However, since this time is unknown, the similarity calculated at the time of the fall does not determine the parent body, but it is considered an acceptable indicator. Therefore, the asteroid 2010 CR19 is likely the parental body of the fireball with best results for similarity on the fall moment between the known minor bodies.

\section{Conclusions}
\label{sec:conclusions}

We investigated a fireball event recorded on 2023 February 27 at 18:15:55.77 UT over northern Sweden, using data acquired from the ALIS\_4D optical system, one station of the Swedish Allsky Meteor Network, and a privately owned camera. From this multi-station dataset, we reconstructed the trajectory and velocity. We also derived the $\alpha$ and $\beta$ parameters characterizing the object's atmospheric flight.

This study identifies a strewn field along the boundary between the Kiruna and Gällivare municipalities in northern Sweden, in the vicinity of road E10 and the Kalix river. 

The derived orbit places the fireball's meteoroid within the Apollo family of Near-Earth Objects (NEOs). Long-term backward numerical integrations suggest that the meteoroid remained in the NEO region for an extended period, experiencing repeated close encounters with Earth that may have ultimately led to the collision on 27 February. Among the known minor bodies, asteroid 2010 CR19 provides the best orbital match under our adopted similarity criteria. However, its dynamical history, shaped by repeated close encounters with the Earth, precludes a firm parent-body association.

Our results demonstrate the feasibility of employing auroral imaging systems for quantitative meteor science, particularly in constraining physical properties. This study highlights the capability of the ALIS\_4D system to support the monitoring and analysis of meteor events, especially bright fireballs. A key advantage of using ALIS\_4D lies in its ability to measure absolute photon flux, which enables detailed investigation of the meteoroid ablation process. Additionally, the system’s use of optical filters and high-speed imaging configurations enhances its dynamic range, making it well-suited for capturing high-brightness events.

\section*{Acknowledgements}
Part of the work developing the analysis pipeline was supported by Swedish National Space Agency NRFP4 grant 2020-304 and the EU project RIT (Rymd för innovation och tillväxt/Space for innovation and growth) with the objective of developing the space region in northern Sweden. The Finnish team acknowledges support from the Academy of Finland (project no. 325806, PlanetS), which facilitated the development of the method used in this study. M.G. thanks Adolfo González Rivera (Alhama Academy) for logistical support and acknowledges funding from the Spanish Ministry of Science, Innovation and Universities (project PID2023-151905OB-I00). DCM and BCBC gratefully acknowledge financial support from the Coordenação de Aperfeiçoamento de Pessoal de Nível Superior (CAPES) and the Conselho Nacional de Desenvolvimento Científico e Tecnológico (CNPq), process No. 405349/2025-4. We thank to Tomoe Taki for valuable support during the review process.

\section*{Data Availability}

All data and materials used in this study are publicly available in \citet{repository}. This includes FITS data and processed data necessary to support the findings of this study.




\bibliographystyle{mnras}
\bibliography{references} 




\appendix

\section{Similarity pipeline}
\label{app:a}

Here we detail the pipeline used to compute the similarity metrics. The first step was to compute d-Drummond and d-Nersvorny between the fireball and all known small bodies. We used the fireball orbital elements present in Table~\ref{tab:orb_ele_both}. The orbital elements of Minor Planets (MP) were obtained from the MPCORB.DAT file provided by the Minor Planet Center\footnote{File downloaded from \url{https://minorplanetcenter.net/iau/MPCORB.html} on 11 May 2025.}. We computed the similarity of the fireball with more than 1.4 millions of MP. We sorted the MPs by criteria, then selected the top $N$ candidates. $N$ was chosen considering the computational capability and the close population to avoid unnecessary computations. Thereafter,  we recomputed the similarity considering the orbital elements of these candidates at the time of the fireball epoch. These ephemerides were obtained from the JPL HORIZONS system\footnote{\url{https://ssd.jpl.nasa.gov/horizons/}}. 

For the second time, we computed the minimum d-Drummond and d-Nersvorny for each MP varying the orbital element in the range of the error(Table~\ref{tab:orb_ele_both}). Finally, we sorted by the d-Drummond and d-Nersvorny to identify the best parental body candidates. This pipeline was implemented in Python and is available open source\footnote{\url{https://github.com/danielk333/dasst}}.

\begin{table}
\caption{Top five minor planets with the smallest d-Drummond and d-Nesvorn\'y values, calculated using orbital elements from the MPCORB file (retrieved May 11, 2025), the fireball’s fall epoch, and accounting for the fireball’s orbital uncertainties (Error).}
\begin{tabular}{llll}\hline 
	\multicolumn{4}{c}{Drummond} \\ \hline	
        &MPCORB & Fall Epoch & Error \\ \hline	
$1^{st}$&2010 CR19 & 2020 DY3 & 2010 CR19  \\
$2^ {nd}$&2020 DY3 & 2023 ED & 2019 RN1  \\
$3^ {rd}$&2023 ED & 2010 CR19 & 2014 DH6  \\
$4^ {th}$&2024 DE1 & 2024 DE1 & 2020 DY3 \\
$5^ {th}$&2014 DH6 & 2014 DH6 & 2017 NQ6 
\\ \hline
	\multicolumn{4}{c}{Nersvorný}\\ \hline	
        &MPCORB & Fall Epoch & Error  \\
\\ \hline
$1^{st}$& 2010 CR19 & 2010 CR19 & 2010 CR19 \\
$2^ {nd}$& 2007 FC3 & 2007 FC3 & 2014 DH6 \\
$3^ {rd}$& 2014 DH6 & 2014 DH6 & 2007 FC3 \\
$4^ {th}$& 2017 YC8 & 2017 YC8 & 2017 YC8\\
$5^ {th}$& 2019 TP1 & 2019 TP1 & 2019 TP1\\ \hline

\end{tabular}
\label{tab:app}
\end{table}

Table~\ref{tab:app} shows the position of the best parent body candidates for the fireball in each step of the pipeline. The variation observed in the ranking of the top five candidates highlights the influence of the orbital epoch on the similarity calculations.


\bsp	
\label{lastpage}
\end{document}